 \documentclass[acmlarge,screen]{acmart}

\AtBeginDocument{%
  \providecommand\BibTeX{{%
    \normalfont B\kern-0.5em{\scshape i\kern-0.25em b}\kern-0.8em\TeX}}}

\setcopyright{acmcopyright}
\copyrightyear{2018}
\acmYear{2022}
\acmDOI{10.1145/1122445.1122456}

\acmConference[CHI '22]{CHI '22: ACM CHI Conference on Human Factors in Computing Systems}{April 30--May 6, 2022}{New Orleans, LA}
\acmBooktitle{CHI '22: ACM CHI Conference on Human Factors in Computing Systems,
  April 30--May 6, 2022, New Orleans, LA}
\acmPrice{15.00}
\acmISBN{978-1-4503-XXXX-X/18/06}

\usepackage{comment}

\begin{document}

\title{Approaches to Identify Vulnerabilities to Misinformation: A Research Agenda}

\author{Nattapat Boonprakong}
\email{nboonprakong@student.unimelb.edu.au}
\affiliation{%
  \institution{University of Melbourne}
  \country{Australia}
 }
\author{Benjamin Tag}
\email{benjamin.tag@unimelb.edu.au}
\affiliation{%
  \institution{University of Melbourne}
  \country{Australia}
}
\author{Tilman Dingler}
\email{tilman.dingler@unimelb.edu.au}
\affiliation{%
  \institution{University of Melbourne}
  \country{Australia}
}

\renewcommand{\shortauthors}{N. Boonprakong, B. Tag, T. Dingler}

\begin{abstract}
  Given the prevalence of online misinformation and our scarce cognitive capacity, Internet users have been shown to frequently fall victim to such information. As some studies have investigated psychological factors that make people susceptible to believe or share misinformation, some ongoing research further put these findings into practice by objectively identifying when and which users are vulnerable to misinformation. In this position paper, we highlight two ongoing avenues of research to identify vulnerable users: detecting cognitive biases and exploring misinformation spreaders. We also discuss the potential implications of these objective approaches: discovering more cohorts of vulnerable users and prompting interventions to more effectively address the right group of users. Lastly, we point out two of the understudied contexts for misinformation vulnerability research as opportunities for future research.
  
\end{abstract}



\maketitle

\section{Introduction}
There is a sheer amount of misinformation circulating on the Internet nowadays. Defined as false or misleading information, misinformation misinforms people and bears many of the harmful effects to individuals and society~\cite{lewandowsky2017beyond}. More recently, misinformation has undermined public health guidance about COVID-19~\cite{Roozenbeek2020Suscept}. Through Internet users who consume online media, misinformation has been shown to spread faster than factually accurate information~\cite{Vosoughi2018TheSpread}. This phenomenon, where users believe and share misinformation without taking deliberate considerations, can be attributed to not only the presence of misinformation but also the psychological vulnerabilities of information consumers~\cite{DeFreitas2013Vulnerabilities}. Humans are not always rational~\cite{simon1957behavioral}; therefore, they are prone to make erroneous decisions, like misinterpreting news or sharing unverified information. 

There are two main actions regarding misinformation: believing and sharing. In the literature, research has explored several user-side factors that govern the susceptibility of people to \textit{believe} online misinformation. Cognitive biases--mental shortcuts we use to achieve faster but less deliberate decisions-- have been marked as factors that make people be gullible to false or unverified information~\cite{pantazi2021social}. \citet{martel_reliance_2020} have shown that reliance on emotion indicates higher belief in false information. Some research have shown that political partisans~\cite{Nikolov2021Right} or those who are older than 65 years~\cite{Brashier2020Aging} are prone to believing fake news. \citet{Roozenbeek2020Suscept} found that higher numeracy skills and better trust in science are indicators of lower susceptibility to misinformation.

On the other hand, some research has investigated why people \textit{share} unverified information online. Motivated by ~\cite{difonzo_rumor_2007, pennycook2020fighting}, \citet{Karami2021Profiling} listed five motivational factors in spreading fake news: uncertainty, anxiety, lack of control, relationship enhancement, and social rank. \citet{Laato2020Why} found that trust in online information and information overload are strong indicators of sharing unverified information. \citet{Chen2015Why} suggested that people share misinformation because of social factors, i.e., self-expression and socialization. In addition, \citet{Mihai2020Exposure} found that social engagement metrics (i.e., the number of likes or shares) increased the tendency to share misinformative contents.

Based on the above-mentioned psychological findings, some research have put these findings into practice by building computing systems that objectively identify potential misinformation spreaders based on their online and offline behaviours~\cite{Giachanou2020Role, Karami2021Profiling, sulflow_selective_2019}. Unlike previous approaches that rely on self-report measures, these approaches quantify how people behave with information and produce insights about their specific behaviours regarding to misinformation. Through behavioural and psycho-physiological measures, these approaches can explicitly identify \textit{when} and \textit{which} users are prone to believe or spread misinformation. 

Different individuals have different backgrounds and different levels of cognitive capacity. Therefore, some information consumers may be more prone than others to misinformation. Echoing the findings in \citet{geeng2020social} that different design interventions may be effective for different users, the objective approaches to identify vulnerable users would also provide a a guide on which misinformation intervention should be deployed to which group of users. Ultimately, interventions would be made more effective as they address the right group of users.


In this position paper, we discuss empirical approaches to identifying people who may be susceptible to misinformation: detecting cognitive biases and profiling vulnerable users. We also highlight their implications of identifying more cohorts of vulnerable users and prompting interventions to address the right group of users. Lastly, we elaborate on challenges and future avenues to investigate vulnerabilities to misinformation. We intend to generate a fruitful discussion in this workshop about the need to explore potential victims of misinformation, identify their vulnerabilities, and employ the right interventions to address the right user cohorts.

\section{Approaches to Identifying Vulnerabilities to Misinformation}
Various studies have investigated why people believe and share misinformation on the Internet. However, most of the studies addressed such questions based on self-report measures, e.g., questionnaires, which are prone to many limitations such as subjectiveness. On the other hand, some research has proposed objective approaches to identifying users who may be susceptible to misinformation. Combining prior psychological findings and behavioural measures, these research efforts explicitly identify which and when users tend to believe and share misinformation. In this paper, we discuss two promising approaches: detecting cognitive bias detection and profiling social media users. 

\subsection{Cognitive Bias Detection}
Humans possess limited cognitive capacity and employ cognitive biases as mental shortcuts, which can lead to irrational judgements. One prominent example is ``confirmation bias'' (also known as \textit{selective exposure}) as a tendency to seek solely information to supports one's perspectives or expectations, while ignoring other dissenting information~\cite{Nickerson1998ConfirmationBias}. Taking the advantage of social media platforms that inform users with contents that reinforce their viewpoints, confirmation bias is easily seen on social media platforms as many users tend to consume and spread content items that match their belief without checking its veracity~\cite{Zollo2018Misinformation}. 

Addressing cognitive biases might be a feasible approach to identify when the user is susceptible to misinformation. Methods to detect confirmation bias have traditionally relied on self-reports~\cite{Clay2013Techniques}, for example, asking users about their news consumption habits~\cite{garrett_turn_2013}. Recent approaches, on the other hand, have used behavioral measures to objectively determine confirmation bias (e.g., time spent on reading news articles). In addition, some research employed psycho-physiological signals as objective measures of confirmation bias, like eye tracking~\cite{marquart_selective_2016, sulflow_selective_2019} and electroencephalography~\cite{moravec_fake_2018}.

However, we note that research on cognitive bias detection has been limited. While we have seen some research quantifying confirmation bias, little attention has been paid to investigate the detection of other cognitive biases that also help the spread of misinformation. The bandwagon effect, for example, is a tendency to adapt certain behaviours (e.g., believing a news piece) because many other people are doing the same. There have been some research efforts in the context of information retrieval to detect such phenomena~\cite{Harris2019Detecting}; thus, it is promising to see the quantification of bandwagon effects in the realm of online information consumption.

\subsection{Profiling Vulnerable Users}
Some ongoing research focused on empirical, objective approaches to profiling potential fake news believers and spreaders by using data mining and machine learning algorithms on real-world social media data. \citet{Shu2018Understand} extracted social media user profiles into explicit features (metadata, for example, the number of posts and the number of followers) and implicit features (e.g., age, gender, personality) and used them to infer users who are prone to believing fake news. \citet{Giachanou2020Role} combined linguistic and the Big Five Personality scores in users' tweets to classify Twitter users who spread fake news. \citet{Karami2021Profiling} developed metrics to measure how likely a Twitter user will spread false information. Informed by research in psychology~\cite{difonzo_rumor_2007, pennycook2020fighting}, they quantified each user profile based on the five categories of motivational features: uncertainty, emotions, lack of control, relationship enhancement, and social rank. Furthermore, this problem led to a collective effort in a challenge at the PAN at CLEF 2020 workshop\footnote{\url{https://pan.webis.de/clef20/pan20-web/}} where several machine learning approaches to detect fake news spreaders were proposed~\cite{rangel2020overview}.


\section{Future Implications: Designing for Different Vulnerabilities}
In this section, we discuss future implications from each of the two presented approaches to identifying those who are vulnerable to misinformation.

\subsection{Cognitive Bias Reflection}
The ability to detect cognitive biases implies that we can prompt users to reflect on their susceptibility to misinformation. (or, in the case of confirmation bias, information that confirms their opinion) Furthermore, interventions could be employed to address users who are subject to cognitive biases. On the other hand, we would be able to more accurately place misinformation interventions on contents that induce users' cognitive biases. For instance, interface designers may place a warning message that prompts users awareness (e.g., "This content might reinforce your belief"~\cite{Rieger2021ThisItem}) or a nudge (e.g., recommending alternative contents to balance out the viewpoints~\cite{Thornhill2019Digital}) on online contents/posts/tweets that pose cognitive biases and susceptibility to believe and spread misinformation.

\begin{figure}[h]
  \centering
  \includegraphics[width=0.4\linewidth]{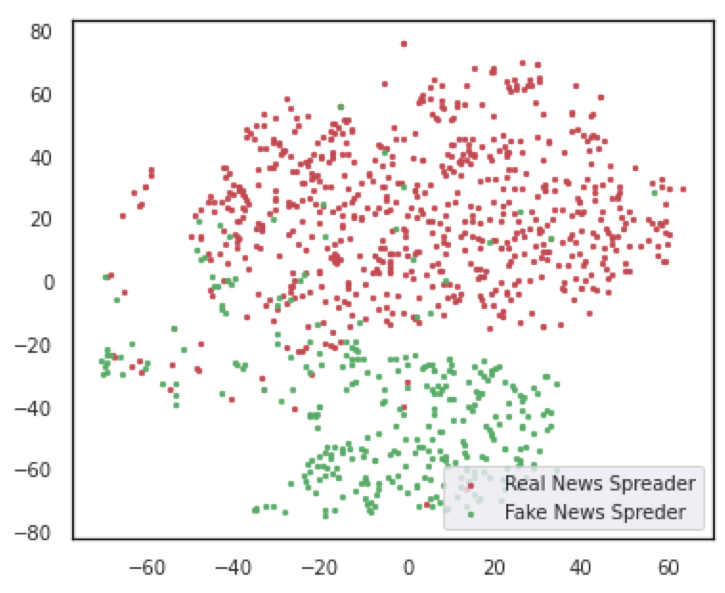}
  \includegraphics[width=0.4\linewidth]{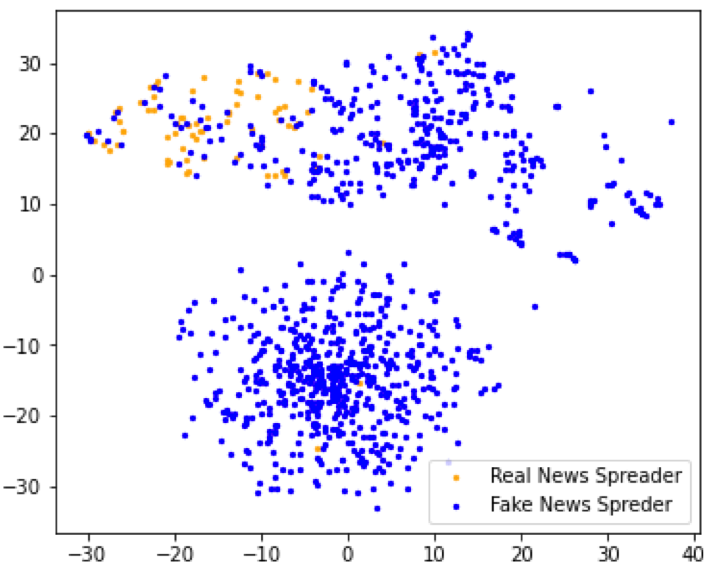}
  \caption{t-SNE visualization of Twitter profiles from the PolitiFact dataset and GossipCop dataset (Obtained from \citet{Karami2021Profiling})}
  \Description{t-SNE visualization of Twitter profiles from the PolitiFact dataset and GossipCop dataset (Obtained from \citet{Karami2021Profiling})}
   \label{t-sne}
\end{figure}

\subsection{Data-driven Identification of Vulnerable Users}
Online behaviour data provide opportunities to model and profile users who may be more susceptible to misinformation. Data-driven approaches could identify more features associated with the users' tendency to spread misinformation online. Linguistic features, for example, might be able to tell the user's age cohort~\cite{Nguyen2013HowOld} or political orientation~\cite{pennacchiotti2011machine}. 

Furthermore, with the abundance of features, we could perform clustering on the feature vectors of user profiles to explore user groups that are represented by clusters of data. Based on the t-SNE visualization of user profiles in \citet{Karami2021Profiling} using PolitiFact and GossipCop datasets~\cite{shu2018fakenewsnet} (see Figure \ref{t-sne}), we may be able to imply that there are some fake news spreaders who exhibited some similarities with those who did not share fake news because their data points are closely located on the representation space. Further investigations would reveal the features of such user groups and proper intervention methods to address them.

The identified vulnerabilities would also inform the design of misinformation interventions that is personalised or targeted according to users' behaviour on platforms. At the same time, misinformation interventions would be made more explainable to the users as interventions are attributed by the vulnerabilities informed by the empirical results.

\section{Challenges: Understudied Context}
Although approaches exist to detect potential consumers and spreaders misinformation, there is more avenues for researchers to investigate vulnerabilities to misinformation. In this section, we discuss two challenges: visual-based media and instant messaging services. Visual information and instant messaging are understudied contexts in misinformation research. Therefore, we argue that more research is needed to address people's vulnerabilities to misinformation in the forms of visuals and private messaging.

\subsection{Visual-based Media}
Visual information (i.e., images and videos) is one of the most prevalent format of misinformation~\cite{Brennen2021Beyond} and it is more likely to persist in our memory~\cite{Defeyter2009Picture}. However, there has been limited research on visual misinformation~\cite{Saltz2021Encounters}, including those in the context of vulnerabilities to believe and disseminate visual misinformation. In the following, We discuss challenges regarding to the two approaches presented in this paper.

In the realm of cognitive bias detection, some research investigated confirmation bias in viewing visual stimuli. \citet{marquart_selective_2016} examined users' selective exposure to political advertisement posters via eye-tracking data. The paper revealed significant effects of political predisposition on time spent looking political ads. From this paper, we argue that gaze data are promising means to identify patterns of visual information consumption. Gaze patterns would reveal how users see, share, and interact with visual misinformation on visual-based social media platforms, like Instagram, Tiktok, Youtube, or Snapchat.

Additionally, there has been little research that investigate potential spreaders of misinformation on visual-based platforms. Yet, \citet{Hussein2020Measuring} investigated YouTube's personalisation algorithm through search engine audits and found that, while demographics (i.e., age, gender, and location) did not affect the amount of misinformation presented, users with watch history on misinformative topics (e.g., 9/11 or flat earth conspiracies) tended to be exposed to more of such videos. We suggest that further investigation is needed to assess the vulnerabilities of conspiracy video consumers. We also argue that visual-based media pose some challenges for user profile investigation. One prominent challenge is that the meaning conveyed by visual information can be highly contextual~\cite{Saltz2021Encounters}. For example, one might share a video containing false claims with accurate user corrections in a caption.

\subsection{Instant Messaging}
Instant messaging (IM) applications (for instance, WhatsApp, WeChat, and Telegram) are one of the most used services on mobile phones. However, such services are fertile grounds for the spread of misinformation and propaganda~\cite{bradshaw2018challenging, garimella2020images}. The role of IM services on misinformation has been understudied and obtaining real-world data has been difficult due to its private, encrypted nature.

We argue that cognitive biases exist in the interactions between users and messages. Users may forward message contents that confirm their beliefs to other users or chat groups. At the same time, some users may belong in one or more chat groups, which consist of people of same interests, e.g., liberal or conservative politics. In such cases, private group chats may be amplification venues for contents that tell predominantly one side of every story. It is therefore interesting to see the detection of cognitive biases in IM settings and interventions that invite users to reflect themselves on, for example, their information diet.

While there has been some studies that explored motivations in sharing fake news on IM platforms~\cite{HerreroDiz2020Teen}, there is little or no research on approaches to objectively identify IM users who may be susceptible to misinformation. Although data on IM platforms are hardly accessible, it is encouraging to see many of the psychological aspects (for example, the pace of communication and emotion expressed as a group) be taken into account when designing systems that detect potential spreaders of misinformation.

\section{Concluding Remarks}
Misinformation does its work in deceiving and misinforming users. However, the question remains how people fall victim of misinformation. Various research efforts have explored psychological factors involved; furthermore, some ongoing research have brought these insights into practice. In this position paper, we discuss two promising approaches to identify people's vulnerabilities towards believing and spreading misinformation. Through cognitive bias detection and user profiling, we explore their implications in (1) prompting misinformation interventions to effectively address the right user groups and (2) identifying more vulnerable user cohorts. We also address existing challenges in identifying users' vulnerabilities in understudied contexts like visual-based media and instant messaging services. In the end, as different users possess different vulnerabilities, we suggest that the ability to identify vulnerable users would allow interventions to effectively address the right group of users.

\bibliographystyle{ACM-Reference-Format}
\bibliography{ws}

\end{document}